\documentclass[11pt]{article}
\usepackage{theorem}
\usepackage{latexsym}
\usepackage{amsmath}

\usepackage[dvips]{graphicx}
\usepackage{graphicx}

\theoremheaderfont{\scshape}
\newtheorem{theorem}{Theorem}[section]
\newtheorem{corollary}{Corollary}[section]
\newtheorem{lemma}{Lemma}[section]
\newtheorem{proposition}{Proposition}[section]
\newtheorem{postulate}{Postulate}[section]
\newtheorem{statement}{Statement}[section]
{\theorembodyfont{\rmfamily}
\newtheorem{example}{Example}[section]}
{\theorembodyfont{\rmfamily}
\newtheorem{remark}{Remark}[section]}
{\theorembodyfont{\rmfamily}
\newtheorem{definition}{Definition}[section]}

\newcommand{\newc}{\newcommand}

\newc{\be}{\begin{equation}}
\newc{\ee}{\end{equation}}
\newc{\bea}{\begin{eqnarray}}
\newc{\eea}{\end{eqnarray}}
\newc{\beas}{\begin{eqnarray*}}
\newc{\eeas}{\end{eqnarray*}}

\newc{\pardt}{\partial_{t}}
\newc{\pardxi}{\partial_{i}}
\newc{\pardts}{\partial_{t^{*}}}
\newc{\pardxis}{\partial_{i^{*}}}
\newc{\pardxj}{\partial_{j}}
\newc{\pardxk}{\partial_{k}}
\newc{\pard}{\partial}

\newc{\s }{\overline}

\newc{\sect}{\section}
\newc{\subs}{\subsection}

\newc{\defi}{\definition}
\newc{\prop}{\proposition}
\newc{\rem}{\remark}
\newc{\lem}{\lemma}
\newc{\exa}{\example}
\newc{\theo}{\theorem}
\newc{\coro}{\corollary}
\newc{\post}{\postulate}
\newc{\state}{\statement}

\begin{document}
\baselineskip0.5cm
\renewcommand {\theequation}{\thesection.\arabic{equation}}
\title{A mathematical model of counterflow superfluid turbulence
describing heat waves and vortex-density waves}

\author{M.
Sciacca$^1$, M.S.~Mongiov\`{\i}$^1$\thanks{Corresponding author.}
and D.~Jou$^2$}

\date{}
\maketitle
\begin{center} {\footnotesize $^1$Dipartimento di Metodi e Modelli Matematici Universit\`a di
Palermo, c/o Facolt\`{a} di Ingegneria,\\ Viale delle Scienze,
90128 Palermo, Italy \\ $^2$ Departament de F\'{\i}sica,
Universitat Aut\`{o}noma de Barcelona, 08193 Bellaterra,
Catalonia, Spain }
\end{center}
\\
\noindent{\bf PACS number(s)}: 67.40.Vs, 67.40.Bz, 47.27.2i,
05.70.Ln
 \footnotetext{E-mail addresses: david.jou@uab.es (D. Jou), mongiovi@unipa.it (M. S. Mongiov\`{\i}),
msciacca@unipa.it (M. Sciacca)}

\begin{abstract}
The interaction between vortex density waves and high-frequency
second sound in counterflow superfluid turbulence is examined,
incorporating diffusive and elastic contributions of the vortex
tangle. The analysis is based on a set of evolution equations for
the energy density, the heat flux, the vortex line density, and
the vortex flux, the latter being considered here as an
independent variable, in contrast to  previous works. The latter
feature is crucial in the transition from diffusive to propagative
behavior of vortex density perturbations, which is necessary to
interpret the details of high-frequency second sound.

\end{abstract}

\section{Introduction}
The study of vortex tangles in superfluid helium II has received
much attention during the last decades, both because of its own
interest and as a first step to understand the classical
turbulence. Many theoretical and experimental studies on
superfluid arguments have enhanced the knowledge on these
intricate phenomena: experimental studies have allowed direct
results and confirmed theoretical intuition, while, on the other
hand, theoretical studies are important not only as a guide for
experiments but also as an explanation of the experimental results
\cite{D2}--\cite{NF-RMP-67-1995}.

It is known that the presence of a heat flow in superfluid helium
II causes the formation of quantized vortex lines, which move
inside the superfluid until a stationary situation is reached, and
whose presence is usually investigated by second sound waves
\cite{D2}--\cite{NF-RMP-67-1995}. But, these waves interact  with
the vortex lines, causing not only a modification of their
velocity and an attenuation of it, but also a modification of the
vortex lines profile and of their motion. In the last years, the
study of non-stationary and inhomogeneous turbulent states has
attracted much attention \cite{MJ1}--\cite{G-PhyB154}. The vortex
lines and their evolution are investigated by second sound waves,
so that it is necessary to analyze in depth their mutual
interactions. In particular, high-frequency second sound may be of
special interest to probe small length scales in the tangle, which
is necessary in order to explore, for instance, the statistical
properties of the vortex loops of several sizes. In fact, the
reduction of the size of space averaging is one of the active
frontiers in second sound techniques applied to turbulence, but at
high-frequencies, the response of the tangle to the second sound
is expected to be qualitatively different than at low frequencies,
as its perturbations may change from diffusive to propagative
behavior \cite{MJ1}--\cite{JMS-Ph.Lett.A}.

In a previous paper \cite{MJ1} a thermodynamical model of
inhomogeneous superfluid turbulence was built up with the aim to
study the mutual interactions between second sound and the vortex
tangle. The fundamental fields of the model were the density
$\rho$, the velocity ${\bf v}$, the internal energy density $E$,
the heat flux $\bf q$ and the average vortex line length per unit
volume $L$. In a successive paper \cite{JMS-Ph.Lett.A}, starting
from this model, a semiquantitative expression for the vortex
diffusion coefficient was obtained and the interaction between
second sound and the tangle in the high-frequency regime   was
studied. In both these works, for the sake of simplicity, the
diffusion flux of vortices ${\bf J}$ was considered as a dependent
variable, collinear with the heat flux $\bf q$, which is
proportional to the counterflow velocity ${\bf V }$, defined as
${\bf V }={\bf v}_n-{\bf v}_s$, ${\bf v}_n$ and ${\bf v}_s$ being
the velocities of the normal and superfluid components,
respectively.

But, in general, this feature is not strictly verified because the
vortices move with a velocity ${\bf v}_L$, which is not collinear
with the counterflow velocity. Indeed, using the vortex filaments
model proposed by Schwarz in
\cite{Schwarz-1982}--\cite{Schwarz-1988}, which describes the
vortex line   by a vectorial function ${\bf s}(\xi,t)$, $\xi$
being the arc-length measured along the curve of the vortex
filament, the velocity of the vortex element is
\cite{D2}--\cite{NF-RMP-67-1995} \be\label{vL} {\bf v}_L={\bf
v}_{sl}+\alpha {\bf s'}\times ({\bf V}-{\bf v_i}) -\alpha'{\bf s'}
\times [{\bf s'} \times ({\bf V}-{\bf v_i})], \ee where $\alpha$
and $\alpha'$ are temperature-dependent friction coefficients
between the normal fluid and the vortex line, ${\bf s'}$ the unit
vector tangent along the vortex line at a given point, and ${\bf
v}_{sl}= {\bf v}_{s}+{\bf v_i}$ the "local superfluid velocity",
sum of the superfluid velocity at large distance from any vortex
line and of the "self-induced velocity", a flow due to all the
other vortices including other parts of the same vortex, induced
by the curvature of all these lines. In (\ref{vL}) the prime
indicates the derivative with respect to the arc-length $\xi$,
that is $s'\equiv \pard s/\pard \xi$. In the "local induction
approximation", the self-induced velocity ${\bf v_i}$ is
approximated by \cite{D2}--\cite{NF-RMP-67-1995}
\be\label{viloc}{\bf v_i}^{(loc)}= \tilde\beta \left[ {\bf
s'}\times {\bf s''} \right]_{s=s_0},\hskip0.3in \hbox{with}
\hskip0.3in \tilde\beta= {\kappa \over 4\pi} \ln\left({c\over a_0
L^{1/2}}\right),\ee
 with $c$ a constant of the order of unity, $\kappa$ the quantum
 of vorticity, given by $\kappa=h/m$, $h$ Planck's constant and
 $m$ the mass of a helium atom,  $a_0$ the size
of the vortex core and ${\bf s''}=\pard^2 s/\pard \xi^2$ the
curvature vector.  The intensity of $\bf v_i$ is $\vert{\bf
v_i}\vert= {\tilde\beta/ R}$, with $R$ the curvature radius of the
vortex line. The coefficient $\tilde\beta$ is linked to the
internal energy per unit length of the vortex line (the tension of
the vortex line) by the relation $\epsilon_V= \rho_s\kappa
\tilde\beta$.

Using the local induction approximation, equation (\ref{vL}) can
be written as  \be {\bf v}_L= (1-\alpha')\tilde\beta   {\bf
s'}\times{\bf s''}
 + \alpha \tilde\beta  {\bf s''}+{\bf v}_{s}
+\alpha {\bf s'}\times {\bf V} -\alpha'{\bf s'} \times ({\bf s'}
\times {\bf V}). \ee Consider now a mesoscopic portion of
turbulent superfluid, contained in a small volume $\Lambda$, which
contains a small vortex tangle. In the following, the vortex
velocity $ <{\bf v}_L>$, averaged in $\Lambda$, will be denoted
with ${\bf v}_{tangle}={\bf v}_{tan}$.

Integrating over the volume $\Lambda$,  recalling that in
counterflow situations, \textrm{i.e.} for $\rho_n {\bf
v}_{n}+\rho_s {\bf v}_{s}=0$, it is ${\bf v}_{s}=-(\rho_n/\rho)
{\bf V}$, and supposing that in $\Lambda$ the counterflow velocity
is constant, one gets  \be {\bf v}_{tan}=(1-\alpha')\tilde\beta <
{\bf s'}\times{\bf s''}>
 + \alpha \tilde\beta < {\bf s''}> -{\rho_n\over \rho}{\bf V}   +
  \alpha <{\bf s'}>\times {\bf V} +\alpha'<{\bf U}-
{\bf s's'} >\cdot  {\bf V}, \ee where $<\cdot>$ stands for the
averaged value of the vector in $\Lambda$.

Suppose that in the small volume $\Lambda$ the vortex tangle is
homogeneous. If the considered volume $\Lambda$ is sufficiently
far from the walls, and therefore does not contain pinned
vortices, one can suppose $<{\bf s'}>=0$ and $<{\bf s''}>=0$,
obtaining \be\label{v_t} {\bf v}_{tan}=<{\bf v}_L>= -{\rho_n\over
\rho}{\bf
V}+\frac{2}{3} \alpha'{\bf \Pi}^s \cdot  {\bf V} +(1-\alpha')\tilde\beta {\bf I} c_1,\ee 
where the tensor ${\bf \Pi}^s=\frac{3}{2}<{\bf U}- {\bf s's'} >$
was studied in \cite{JM-PRB74}, while the vector ${\bf I}= \frac{<
{\bf s'}\times{\bf s''}>}{< |{\bf s''}|>}$ and the scalar
$c_1=\frac{< |{\bf s''}|>}{\Lambda L^{3/2}}$ were introduced by
Schwarz \cite{Schwarz-1988} in a microscopic approach to the
dynamics of superfluid vortex tangles. From (\ref{v_t}) one sees
that ${\bf v}_{tan}\parallel {\bf V}$ only when ${\bf \Pi}^s={\bf
U}$, \textrm{i.e.} when the distribution of ${\bf s'}$ in the
tangle is isotropic, and the anisotropy vector ${\bf I}$ is
collinear to the counterflow velocity ${\bf V}$. However,
experiments and numerical simulations show that these hypotheses
are only approximately verified.

In the hypothesis of isotropy of the tangle (${\bf \Pi}^s={\bf
U}$), the assumption ${\bf v}_{tan}=0$ implies ${\bf I}
\parallel {\bf V}$. But, in general, one could have ${\bf v}_{tan}\neq 0$ and  ${\bf
I}$ not collinear to ${\bf V}$, which means ${\bf v}_{tan}=\s A
{\bf V} +\s B {\bf I},$ with $\s A$ and $\s B$ suitable
coefficients coming from the relation (\ref{v_t}). Therefore, the
hypothesis ${\bf v}_{tan}$ collinear with ${\bf V}$ is not in
exact agreement with the microscopic results of the vortex
filament model. For this reason, the aim of this paper is to build
up a model of inhomogeneous counterflow superfluid turbulence, in
which the flux of the vortex lines, which is parallel to ${\bf
v}_{tan},$ is taken as an additional independent variable.

In Section 2 a relaxational generalization of the diffusion
equation for vortex lines is proposed, analogous to the well known
Maxwell-Cattaneo generalization of Fick, Fourier, Ohm and
Newton-Stokes classical transport laws \cite{JCL-book-EIT,
MR-libro}. There, the corresponding generalization of the entropy
in order to achieve compatibility between the relaxational
transport laws and the second law of thermodynamics in a
simplified model is studied, in which only the line density $L$
and its diffusion flux ${\bf J}$ are fundamental variables. This
simplified model allows us to understand the contribution of ${\bf
J}$ to the nonequilibrium entropy but does not describe the
interaction between vortex density waves and second sound. For
this reason, in Section 3 a more general analysis, including as
independent variables of the theory not only the line density $L$
and the diffusion flux of vortices, but also the energy and the
heat flux is undertaken. The mathematical formalism is  physically
motivated to explore the interactions between heat waves and
vortex-density waves. In Section 4, the physical meaning of the
several coefficients appearing in the model are examined, and in
Section 5  wave propagation in this more complete model in
uncoupled and coupled situations is studied and the results are
compared with those obtained in \cite{MJ1,JMS-Ph.Lett.A}.

\section{Simple formulation of vortex-density waves}
\setcounter{equation}{0} This Section aims to provide a simplified
view of the behavior of turbulent vortices in inhomogeneous
situations, with special emphasis on the transition from diffusive
behavior at low frequencies to propagative behavior at high
frequencies. This example provides the physical motivation for the
wider treatment presented in Section 3.

In this Section the constitutive equation for the diffusive flux
of vortex lines is generalized, by including relaxational effects
due to their inertia. Thus, one will be interested in the
evolution of inhomogeneous vortex tangles or in the response to
external perturbations inducing  local changes in the vortex line
density. Usually, in counterflow situation (i.e. under vanishing
barycentric velocity) one considers homogeneous vortex tangles and
the evolution equation of $L$ is assumed to be the well-known
Vinen's equation \cite{D2}--\cite{NF-RMP-67-1995},
\cite{Vinen-PRSL-1957} \be\label{dL-su-dt1} \frac{d L}{d t} =
\alpha_v V L^{3/2} - \beta_v \kappa L^2\equiv\sigma_L, \ee where
$\sigma_L$ stands for the net vortex production per unit volume
and time and $\alpha_v$ and $\beta_v$ are numerical coefficients.

When one assumes inhomogeneous vortex tangles, with $L$ differing
from point to point, a further contribution should be added to
(\ref{dL-su-dt1}), thus becoming

\be\label{dL-su-dt2} \frac{d L}{d t}= - \nabla\cdot{\bf
J}+\sigma_L. \ee In (\ref{dL-su-dt2}), it appears a new quantity,
${\bf J}$, the flux of vortex lines. In principle, one could
expect that ${\bf J}$ will be related to the gradient of $L$, in
analogy with the well-known Fick's law for matter diffusion. In
fact, in \cite{MJ1}  a thermodynamic formalism leading to such a
transport equation is studied. In some occasions, however,
inertial effects may be relevant, and they will contribute to the
constitutive law for ${\bf J}$ with a relaxation term.

The aim of this Section is to write an evolution equation for
${\bf J}$ which is compatible with the second law of
thermodynamics. The dependence of the constitutive relations on
the temperature $T$ and on the heat flux ${\bf q}$ will be
neglected. In the following Section this simplified hypothesis
will be abandoned. To achieve a consistent thermodynamic
formalism, with a positive definite entropy production, the
entropy must be extended by including ${\bf J}$ in the set of
independent variables, as in extended thermodynamics. The
corresponding generalized Gibbs equation is \cite{JCL-book-EIT}

\be\label{ro-ds} \rho ds =- T^{-1}\mu^L dL- T^{-1}\tilde{\alpha}{\bf
J}\cdot d{\bf J}, \ee
 where $s$ is the entropy per unit mass, $\tilde{\alpha}$   a coefficient that will be identified
below, and $\mu^L$   the vortex chemical potential. Equation
(\ref{ro-ds}) can be written in terms of time derivatives as

\be\label{ro-spunto1}\rho \dot s = -T^{-1}\mu^L \dot L-
T^{-1}\tilde{\alpha}{\bf J}\cdot \dot {\bf J}. \ee Substituting
(\ref{dL-su-dt2}) in (\ref{ro-spunto1})  one has

\be\label{ro-spunto2}\rho \dot s +\nabla\cdot\left(-T^{-1}\mu^L {\bf
J}\right) =
 {\bf J}\cdot\left[-\nabla(T^{-1}\mu^L) -T^{-1}\tilde{\alpha} \dot {\bf J} \right]
 -T^{-1}\mu^L\sigma_L, \ee
where $-T^{-1}\mu^L {\bf J}$  may be interpreted as the vortex
contribution to the entropy flux, and the term on the right-hand
side as the entropy production. To ensure the positive character
of the diffusion contribution to the latter, one writes

\be\label{JL1} {\bf J} =\gamma\left[-\nabla(T^{-1}\mu^L)
-T^{-1}\tilde{\alpha} \dot {\bf J}
  \right], \ee
with $\gamma$ a positive phenomenological coefficient. A
relaxation time $\tau_J$ may be defined as $\tau_J=\gamma
T^{-1}\tilde{\alpha}$. With this identification of the relaxation
time, (\ref{JL1}) may be written as

\be\label{tauJpunto} \tau_J \dot{{\bf J}}+{\bf J}=-D\nabla L, \ee
with $D$ a vortex diffusion coefficient identified as $D=\gamma
T^{-1} \frac{\pard \mu^L}{\pard L}$. By using dimensional
analysis, $\tau_J$ can be taken of the form $\tau_J=(\gamma_1
\kappa L)^{-1}$, where $\gamma_1$ is a dimensionless
phenomenological coefficient. For fast variations of ${\bf J}$
--- namely, in high-frequency experiments --- the first term on the
left-hand side of (\ref{tauJpunto}) becomes dominant.

Combination of (\ref{tauJpunto}) with (\ref{dL-su-dt2}),
neglecting second-order terms in $\nabla L$, yields

\be\label{tauL2punti} \tau_J \ddot{L}+\dot{L}=D\nabla^2 L+\tau_J
\dot{\sigma}_L+\sigma_L.\ee For low values of the frequency, the
first term of the equation (\ref{tauJpunto}) is negligible and one
has a reaction-diffusion equation, whereas if the frequency is
high, the first term is dominant and (\ref{tauL2punti}) yields a
wave equation for $L$.

This brief presentation, based on the simplest version of the
so-called Extended Thermodynamics \cite{JCL-book-EIT}, has been
enough to give us a simple introduction to the topic studied
below, namely, the transition from a diffusive behavior to an
undulatory behavior of the vortex tangle and to the new physical
features related with ${\bf J}$ as an independent variable.
However,  the evolution equation for ${\bf J}$ is introduced as an
additional equation to the previous system proposed in \cite{MJ1},
ignoring the couplings with other possible phenomena, mainly,
high-frequency second sound. Indeed, in this case, the vortex
tangle will not behave as a diffusive system but as an elastic
matrix, and the dispersion relation for second sound will be
changed with respect to the previous model studied in \cite{MJ1}.
It is need to know in detail these changes in the dispersion
relation, because of the instrumental importance of second sound
as a probe for the vortex tangle.

\section{Balance equations and constitutive theory}
\setcounter{equation}{0}

To deal with sufficient generality with the interactions between
second sound and the dynamics of the vortex tangle, one builds up
a thermodynamical model of inhomogeneous counterflow superfluid
turbulence, which chooses as fundamental fields the energy density
$E$, the heat flux $\bf q$, the averaged vortex line length per
unit volume $L,$ and the vortex diffusion flux ${\bf J}$. Because
experiments in counterflow superfluid turbulence in the linear
regime are characterized by a zero value of the barycentric
velocity ${\bf v}$, in this paper one  does not consider ${\bf v}$
as independent variable. In a more complete model ${\bf v},$ and
$\rho$ will be also fundamental fields.

It is known that the heat flux ${\bf q}$ is linked to the
counterflow ${\bf V}$ through the relation ${\bf q} = \rho_s T s
{\bf V},$ and here one prefers choosing ${\bf q}$ as variable
because it is the macroscopic variable appearing in the balance
equation for the energy, and it may be controlled in the
experiments.

Consider the following balance equations

\be\label{siste-balance} \left\{
  \begin{array}{l}
    \pardt E  +
 \pardxk q_k =0 \\
    \pardt q_i  + \pardxk J^q_{ik}  =\sigma_i^q \\
    \pardt L + \pardxk J_k =\sigma_L \\
    \pardt J_i  +\pardxk F_{ik} =\sigma_i^{{\bf J}}
  \end{array}
\right. \ee where $\pardt$ stands for $\pard/\pard t$ and
$\pardxk$ for $\pard/\pard x_k$, $E$ is the specific energy per
unit volume of the superfluid component plus the normal component
plus the vortex lines, $J^q_{ij}$ the flux of the heat flux, $J_i$
the flux of vortex lines, and $F_{ij}$ the flux of the flux of
vortex lines; $\sigma_{i}^q$, $\sigma_L$ and $\sigma_i^{{\bf J}}$
are the respective production terms.  Since in this work one is
interested to study the linear propagation of the second sound and
vortex waves, the convective terms have been neglected.

If one supposes that the fluid is isotropic, the constitutive
equations for the fluxes $J^q_{ij}$ and $F_{ij}$, to the first
order in $q_i$ and $J_i$, can be expressed in the form

\be\label{siste-costitutive}
\begin{array}{c}
  J_{ik}^q=\beta (E,L)\delta_{ik}, \\
  F_{ik}=\psi (E,L)\delta_{ik}.
\end{array}\ee
 Restrictions on these relations are obtained
imposing the validity of the second law of thermodynamics,
applying Liu's procedure \cite{MR-libro, Liu}. This method
requires the existence of a scalar function $S$ and a vector
function $J_k^S$ of the fundamental fields, namely the entropy per
unit volume and the entropy flux per unit volume respectively,
such that the following inequality

\be\label{entropi-magg}
\begin{array}{c}
  \pardt S +\pardxk J_k^S -\Lambda^E \left[ \pardt E+
\pardxk q_k \right]-\Lambda_{i}^q
 \left[\pardt q_i +\pardxk J^q_{ik}-\sigma_i^q \right] \\
\\
  \hskip0.2in
-\Lambda^L \left[ \pardt L + \pardxk J_k-\sigma_L
 \right]-\Lambda_{i}^{J} \left[\pardt J_i +
\pardxk F_{ik}-\sigma_i^{J}  \right]
 \ge 0 ,
\end{array}\ee
is satisfied for arbitrary fields $E$, $q_i$, $L$ and $J_i$. In
this inequality, which expresses the second law of thermodynamics,
$S$ and $J_k^S$ are objective functions of the fundamental fields.
 In order to make the theory internally consistent, one must
consider for  $S$ and $J_k^S$ approximate constitutive relations
to second order in $q_i$ and $J_i$ \be\label{entr-flussoentr}
S=S_0(E,L)+S_1(E,L)q^2+S_2(E,L)J^2+S_3(E,L)q_iJ_i, \hskip0.2in
J_k^s=\phi^q(E,L) q_k+\phi^J(E,L) J_k . \ee

 The quantities $\Lambda^E$, $\Lambda_i^q$, $\Lambda^L$ and
$\Lambda_{i}^{J}$ are Lagrange multipliers, which are also
objective functions of $E$, $q_i$, $L$ and $J_i$; in particular,
one puts
\[
\Lambda^E=\Lambda^E(E,L,q_i,J_i)=\Lambda^E_0(E,L)+\Lambda^E_1(E,L)q^2+\Lambda^E_2(E,L)J^2+\Lambda^E_3(E,L)q_iJ_i,
\]
\[
\Lambda^L=\Lambda^L(E,L,q_i,J_i)=\Lambda^L_0(E,L)+\Lambda^L_1(E,L)q^2+\Lambda^L_2(E,L)J^2+\Lambda^L_3(E,L)q_iJ_i,
\]

\be\label{Lamq-LamJ} \Lambda_i^q= \lambda_{11}q_i+ \lambda_{12} J_i
\quad \textrm{and} \quad \Lambda_{i}^{J} = \lambda_{21}q_i+
\lambda_{22} J_i ,\ee
 with $\lambda_{m n}=\lambda_{m n}(E,L)$.
 The constitutive theory is obtained imposing in (\ref{entropi-magg}) that the coefficients of all
derivatives vanish. Imposing that the coefficients of the  time
derivatives are zero, one obtains

\be\label{deEntropia} dS =\Lambda^E dE+\Lambda_i^q ~d
q_i+\Lambda^L dL+ \Lambda_i^J dJ_i. \ee Note that $S=\rho s,$
therefore this equation generalizes equation (\ref{ro-ds}) when
energy and heat flux variations are taken into account. In the
same way, imposing that the coefficients of space derivatives
vanish, one finds

\be\label{deFlusso-Entropia} dJ_k^S= \Lambda^E d q_k+ \Lambda_i^q d
{J}_{ik}^q + \Lambda^L d J_k + \Lambda_i^J d F_{ik} . \ee

 Substituting now (\ref{siste-costitutive}), (\ref{entr-flussoentr}) and (\ref{Lamq-LamJ}) in (\ref{deEntropia}-\ref{deFlusso-Entropia}),
 one gets

\be\label{s1-s2-s3} S_1={1\over 2}\lambda_{11},\hskip0.3in
S_2={1\over 2}\lambda_{22},\hskip0.3in
S_3=\lambda_{12}=\lambda_{21}, \ee

\be\label{phi1-phi2}\phi^q=\Lambda^E_0 ,   \hskip0.6in
\phi^J=\Lambda^L_0 , \ee

\be\label{des0-des1}dS_0 = \Lambda_0^E d E+\Lambda_0^LdL ,
\hskip0.3in dS_1 = \Lambda_1^E d E+\Lambda_1^LdL, \ee

\be\label{des2-des3} dS_2 = \Lambda_2^E d E+\Lambda_2^LdL,
\hskip0.3in
 dS_3 = \Lambda_3^E d E+\Lambda_3^LdL, \ee

\be\label{dephi1-dephi2}d\phi^q=\lambda_{11} d \beta + \lambda_{21}
d \psi , \hskip0.5in
 d\phi^J =\lambda_{12} d \beta + \lambda_{22} d \psi .  \ee

In particular, one obtains to the second order in ${\bf q}$ and
${\bf J}$ the following expressions the entropy and for the
entropy flux

\be\label{entr-flussoentr-conlam}S=S_0 +{1\over
2}\lambda_{11}q^2+{1\over 2}\lambda_{22} J^2+
 \lambda_{12}q_iJ_i ,
 \hskip0.2in
J_k^s=\Lambda^E_0 q_k+\Lambda^L_0 J_k . \ee
 It remains the following residual inequality for the entropy production

\be\label{sigmaS}\sigma^S=\Lambda_i^q \sigma_i^q  +\Lambda^L
\sigma_L+\Lambda_i^J
 \sigma_i^J \ge 0. \ee

Now,  the relations obtained  are analyzed in detail. One first
introduces a generalized temperature as the reciprocal of the
first-order part of the Lagrange multiplier of the energy

\be\label{Lam0}\Lambda_0^E = \left[{\partial S_0\over \partial
E}\right]_L = {1\over T}. \ee
 Observe that, in the laminar regime (when $L=0$), $\Lambda_0^E$
 reduces to the inverse of the absolute temperature of thermostatics. In the
 presence of a vortex tangle the quantity (\ref{Lam0}) depends also on
 the line density $L$.

As in \cite{MJ1}, writing equation  $(\ref{des0-des1})_1$ as

\be\label{deE}dS_0= \frac{1}{T} dE +\Lambda_0^L dL= \frac{1}{T} dE
-\frac{\mu_0^L}{T} dL,\ee one can identify the quantity
$-{\Lambda_0^L/\Lambda_0^E}=-T\Lambda_0^L$ with the chemical
potential of vortex lines (near equilibrium)

\be\label{TLam0} -T \Lambda_0^L = \mu^L. \ee

From (\ref{deE}) one obtains the integrability condition

\be \frac{\pard E}{\pard L} =T^2 \frac{\pard}{\pard
T}\left(-\frac{\mu_0^L}{T}\right).\ee

Neglecting in (\ref{deEntropia}) second order terms in ${\bf q}$
 and ${\bf J}$, and using relations (\ref{s1-s2-s3}), (\ref{Lam0})
 and (\ref{TLam0}), the following expression for the entropy
 density $S$ is obtained

 \be\label{deEntropia2}
 d S=\frac{1}{T}d E-\frac{\mu^L}{T}dL+\lambda_{11} q_i d q_i+\lambda_{22} J_i d J_i
 +\lambda_{12} ( J_i d q_i+q_i d J_i). \ee

Consider now equations (\ref{dephi1-dephi2}), which one rewrites
using (\ref{phi1-phi2}) and (\ref{TLam0}) as

\be\label{de-1suT} d\left({1\over T}\right)=\lambda_{11} d \beta +
\lambda_{21} d \psi ,
  \hskip0.5in
 d\left(-{\mu^L\over T}\right) =\lambda_{12} d \beta + \lambda_{22} d \psi .
 \ee
From these equations, one obtains the following relations

\be\label{debeta}d\beta ={\lambda_{22}\over
\lambda_{11}\lambda_{22}-{\lambda_{12}}^2} d\left({1\over
T}\right)+{\lambda_{12}\over
\lambda_{11}\lambda_{22}-{\lambda_{12}}^2}
 d\left({\mu^L\over T}\right),
\ee

\be\label{depsi}d \psi = -{\lambda_{12}\over
\lambda_{11}\lambda_{22}-{\lambda_{12}}^2} d\left({1\over
T}\right)-{\lambda_{11}\over
\lambda_{11}\lambda_{22}-{\lambda_{12}}^2}
 d\left({\mu^L\over T}\right),\ee
from which, putting

\be\label{debeta-sudeTeL-depsi-pure}{\partial \beta\over\partial
T}=\xi ,  \hskip0.3in
 {\partial \beta\over\partial L}= \chi ,  \hskip0.3in
 {\partial \psi\over\partial T}= \eta ,  \hskip0.3in
 {\partial\psi\over\partial L}= \nu  ,    \ee
one obtains

\be\label{lam1chi-lam1zet}\lambda_{11}\chi+\lambda_{21}\nu =0,
\hskip0.5in \lambda_{11}\xi +\lambda_{21} \eta =-{1\over T^2}, \ee

\be\label{lam3zet-lam3chi}\lambda_{12}\xi+\lambda_{22}\eta
={\partial\over\partial T}\left(-{\mu_0^L\over T}\right),
 \hskip0.3in \lambda_{12}\chi
+\lambda_{22} \nu ={\partial\over\partial L}\left(-{\mu_0^L\over
T}\right), \ee and also

\be\label{zet-eta}\xi={1\over N}\left[ -{1\over T^2}\lambda_{22}+
\lambda_{12}{\partial \over\partial T}\left({\mu_0^L\over
T}\right)\right], \hskip0.1in \eta={1\over N}\left[ {1\over
T^2}\lambda_{12}- \lambda_{11}{\partial \over\partial
T}\left({\mu_0^L\over T}\right)\right], \ee

\be\label{chi-nu}\chi={1\over T}{\lambda_{12}\over N} {\partial
\mu_0^L\over\partial L}, \hskip0.4in \nu=-{1\over
T}{\lambda_{11}\over N} {\partial \mu_0^L\over\partial L}, \ee
where $N=\lambda_{11}\lambda_{22}-{\lambda_{12}}^2,$ and $\nu$ is
a positive coefficient because it is the square of the velocity of
the vortex wave, as it will be shown in the next section.

Finally, one obtains for the entropy flux

\be\label{Fluss-entro} J_k^s={1\over T} q_k-{\mu_0^L\over T} J_k,
\ee which is analogous to the usual expression of the entropy flux
in the presence of a mass flux and heat flux, but with the second
term related to vortex transport rather than to mass transport.

Observe that the expression of the entropy flux
(\ref{Fluss-entro}) obtained in this Section is in agreement with
  (\ref{ro-spunto2}) when the dependence on the
heat flux is neglected. In the same way, comparing the expression
of the entropy (\ref{deEntropia2}) with the generalized Gibbs
equation (\ref{ro-ds}), proposed in the simplified model in
Section~2, and keeping in mind equation (\ref{JL1}) one gets
\[
\lambda_{22}=-\tilde \alpha T^{-1}=-\frac{\tau_J}{\gamma},
\]
thus furnishing a physical meaning of the coefficient
$\lambda_{22}$ appearing in previous equations.

Finally, substituting  the  constitutive equations
(\ref{siste-costitutive})  in system (\ref{siste-balance}), and
using the relations (\ref{debeta}-\ref{chi-nu}), the following
system of field equations is obtained

\be\label{siste-Ep-qp-Lp-Jp} \left\{
  \begin{array}{ll}
   \pardt E +\pardxj q_j=0 \\
    \pardt q_i+ \xi \pardxi T+\chi \pardxi L  =\sigma^q_i\\
    \pardt L +\pardxj J_j =\sigma^L\\
    \pardt J_i+ \eta \pardxi T +\nu \pardxi L =\sigma^J_i.
  \end{array}
\right.  \ee The coefficients $\gamma$ and $\eta$ describe cross
effects linking the dynamics of $\bf q$ and $\bf J$ with $L$ and
$T$, respectively. Thus, they are expected to settle an
interaction between heat waves and vortex waves, whose study is
one of the aims of the present work.  The production terms
$\sigma$ must also be specified. Regarding $\sigma^q_i$, since
only counterflow situation is considering , a simplified
expression, already noted in literature \cite{JLM,M-PRB48}, is
assumed

\be\label{prod-term-q} \vec{\sigma}^{\bf q}=-\frac{1}{3} \kappa
B_{HV} L {\bf \Pi}^s \cdot {\bf q}, \ee where
$B_{HV}=\frac{2\rho}{\rho_n}\alpha$ is the Hall-Vinen coefficient
\cite{D2} and ${\bf \Pi}^s=\frac{3}{2}<{\bf U}- {\bf s's'} >$ is
the symmetric tensor mentioned in Section 1. If one assumes
isotropy in the plane $yz$, this tensor ${\bf \Pi}^s$ can be
written as \cite{JM-PRB74,PS-PB2007} \be\label{tensorPi} {\bf
\Pi}^s=\frac{3}{2}\left(
                         \begin{array}{ccc}
                           2a & 0 & 0 \\
                           0 & 1-a & 0 \\
                           0 & 0 & 1-a \\
                         \end{array}
                       \right),
\ee where $0\leq a \leq \frac{1}{3}$ is a parameter characterizing
the anisotropy of the tangle  such  that $<s_y'^2>=<s_z'^2>=a$ and
$<s_x'^2>=1-2a$. If the tangle is completely anisotropic, as in
the case of a regular array produced by rotation, then $a=0$,
whereas if it is isotropic then $a=\frac{1}{3}$. This term
describes a friction force when ${\bf q}$ is orthogonal to vortex
lines and null force when it is parallel to them. A further
dissipative term, proportional to the binormal vector ${\bf I}$
introduced in (\ref{v_t}), could be added to right-hand side in
(\ref{prod-term-q}) (see the Appendix of Ref.~\cite{MJ1}), but it
is neglected here because its contribute is small compared to the
right-hand side of (\ref{prod-term-q}). For the production term
$\sigma^L$, one chooses the Vinen's production and destruction
terms, equation (\ref{dL-su-dt1}), which one can write in terms of
$L$ and of the absolute value of $q$ using the relation ${\bf
q}=\rho_s T s {\bf V}$

\be\label{prod-term-L} \sigma^L=-B L^2+A q L^{3/2}, \ee where
$A=\alpha_v/\rho_s T s$ and $B=\beta_v \kappa$. For the production
term of vortex line diffusion, one assumes the following
relaxational expression (see the relation below equation
(\ref{tauJpunto}))

\be\label{prod-term-J} \vec{\sigma}^{J}=-\gamma_1 \kappa L {\bf
J}=- \frac{{\bf J}}{\tau_J}, \ee where the positive coefficient
$\gamma_1$ can depend on the temperature $T$; with this
expression, in isothermal situations, one would have a diffusion
coefficient given by $D=\tau_J \nu = -\frac{\tau_J}{T
\lambda_{22}} \frac{\pard \mu^L}{\pard L}$. Note that in
(\ref{prod-term-q}) and (\ref{prod-term-J}) one has assumed that
the respective production terms of ${\bf q}$ and ${\bf J}$ depend
on ${\bf q}$ and ${\bf J}$, respectively, but not on both
variables. In more general terms, one could assume that both
production terms depend on the two fields ${\bf q}$ and ${\bf J}$
simultaneously.

Analyzes, now, the entropy production (\ref{sigmaS}) which, with
the expressions of the production terms defined above, becomes
\be\label{prod-entr} \sigma^S = -\lambda_{11} \varpi \left({\bf
\Pi}^s \cdot {\bf q}\right)_i q_i -\lambda_{12} \left(\varpi
\left({\bf \Pi}^s \cdot {\bf q}\right)_i+\gamma_1 \kappa
q_i\right) J_i-\lambda_{22} \gamma_1 \kappa
J_i^2-\frac{\mu^L}{T}\left(-B L+A q L^{1/2}\right)\geq 0, \ee
where $\varpi=\frac{1}{3}\kappa B_{HV}$ and $\gamma$ is the
positive phenomenological constant defined in (\ref{JL1}). Looking
at the expression (\ref{prod-entr}), one notes that the entropy
production  $\sigma^S$ is positive when a suitable choice of the
coefficients $\lambda_{11}$ and $\lambda_{12}$ is made. In
particular, assuming that ${\bf \Pi}^s={\bf U}$, $\sigma^S$ in
(\ref{prod-entr}) is a quadratic form on the variables $|q|$,
$L^{1/2}$ and $|J|.$ Therefore, (\ref{prod-entr}) is verified if
the matrix

\be \left(
                         \begin{array}{ccc}
                           -\lambda_{11} \varpi & -\frac{1}{2T}\mu^L A & -\frac{1}{2} \lambda_{12}(\varpi+\gamma_1 \kappa) \\
                           -\frac{1}{2T}\mu^L A &  \frac{1}{T}\mu^L B & 0 \\
                           -\frac{1}{2} \lambda_{12}(\varpi+\gamma_1 \kappa) & 0 & - \lambda_{22} \gamma_1 \kappa  \\
                         \end{array}
                       \right) \ee
is semidefinite positive. This implies that the coefficient
$\lambda_{11}$ has to be negative and, being $\pard \mu_0^L /\pard
L >0$, from the relation (\ref{chi-nu}) one deduces that $N$ is
positive.

Observe that the field equations for $\bf q$ and $\bf J$ can be
written also as

\be\label{deqi-su-dt}{\partial q_i\over \partial t}
-{\lambda_{22}\over NT^2} {\partial T \over\partial x_i} +
{\lambda_{21}\over N } {\partial
 \over \partial x_i}\left({\mu^L\over T}\right)  =-\frac{1}{2} \kappa B_{HV} L \left[(3a-1)q_1 \delta_{1i}+(1-a)q_i \right], \ee

\be\label{deJL-su-dt}{\partial J_i\over
\partial t}  + {\lambda_{12}\over NT^2}{\partial T \over
\partial x_i}-{\lambda_{11}\over N }{\partial   \over \partial x_i}
\left({\mu^L\over T}\right) =-\gamma_1 \kappa L J_i. \ee Comparing
the equation (\ref{deJL-su-dt}) with (\ref{JL1}), one deduces that
they can be identified with each other if one puts
$\lambda_{12}=0,$ $\lambda_{21}=0$, $\lambda_{22}=-
T^{-1}\tilde{\alpha}$ and $\gamma_1^{-1}=T^{-1}\gamma
\tilde{\alpha} \kappa L .$ Observe also that under this hypothesis
equation (\ref{deqi-su-dt}) becomes

\[
{\partial q_i\over \partial t}+\xi {\partial T\over \partial
x_i}=\sigma^q_i,
\]
where $\xi=-\frac{1}{\lambda_{11}T^2}$. This latter equation is
identical to that used in Refs.~\cite{JLM,M-PRB48}, where the
fields $L$ and $J_i$ were considered as dependent variables.

\section{Physical meaning of the coefficients of proposed equations}
\setcounter{equation}{0} In order to determine the physical
meaning of the coefficients appearing in equations
(\ref{siste-Ep-qp-Lp-Jp})--(\ref{prod-term-J}), concentrate first
the attention on the equations for $L$ and $\bf J$,

\be\label{deLdet} \pardt L+\pardxi J_i=\sigma^L\ee
\be\label{deJidet} \pardt J_i+\eta \pardxi T+\nu \pardxi L=
\sigma^{J_i}=-\gamma_1 \kappa L J_i. \ee Supposing that ${\bf J}$
varies very slowly, Eq. (\ref{deJidet}) gets the form

\be {\bf J}=-\frac{\eta }{\gamma_1\kappa L}\nabla T-\frac{\nu
}{\gamma_1\kappa L}\nabla L.\ee

Substituting it in (\ref{deLdet}), one obtains \be\label{deLdet5}
\pardt L=\frac{\eta }{\gamma_1\kappa L}\nabla^2 T+\frac{\nu
}{\gamma_1\kappa L}\nabla^2 L+\sigma^L.\ee It is then seen that
the coefficient $\frac{\nu }{\gamma_1\kappa L}\equiv D_1$
represents the diffusion coefficient of vortices as already
introduced in (\ref{tauL2punti}) in a simpler setting. Coefficient
$\frac{\eta }{\gamma_1\kappa L}\equiv D_2$ may be interpreted as a
thermodiffusion coefficient of vortices because it links the
temperature gradient to vortex diffusion. In other terms, this
implies a drift of the vortex tangle. Detailed measurements have
indeed shown [1, pag.216] a slow drift of the tangle towards the
heater; this indicates that $\eta<0$ and small. The hypothesis
$\eta=0$ corresponds to $D_2=0$, i.e. the vortices do not diffuse
in response to a temperature gradient. Now, focus the attention on
the equations of $T$ and $\bf q$

\be\label{rocv5} \rho c_V\pardt T+\rho \epsilon_L \pardt L+\pardxi
q_i=0,\ee \be\label{detqi5} \pardt q_i+\xi \pardxi T+\chi \pardxi
L=\vec{\sigma}^{\bf q}=-\frac{1}{3} \kappa B_{HV} L {\bf \Pi}^s
\cdot {\bf q}. \ee

Supposing $\pardt q_i$ negligible in (\ref{detqi5}), one gets \be
\left({\bf \Pi}^s \cdot {\bf q}\right)_i=-\frac{3\xi}{\kappa
B_{HV} L}\nabla T-\frac{3\chi}{\kappa B_{HV} L}\nabla L\ee that is
\be\label{q_i} q_i=-\frac{3\xi}{\kappa B_{HV} L}\left({\bf
\Pi}^s\right)^{-1}\nabla T-\frac{3\chi}{\kappa B_{HV} L}\left({\bf
\Pi}^s\right)^{-1}\nabla L.\ee The first term in (\ref{q_i}) may
be identified as a tensorial thermal diffusivity, and the second
one is analogous to Soret diffusion term, which describes a
coupling between heat flux and concentration gradient in usual
fluids mixtures; here, instead of the concentration of a chemical
species, one has a vortex density gradient.

Substituting (\ref{q_i}) in (\ref{rocv5}), one gets \be\rho
c_V\pardt T+\rho \epsilon_L \pardt L=-\frac{3\xi}{\kappa B_{HV}
L}\left({\bf \Pi}^s\right)^{-1}\nabla^2 T-\frac{3\chi}{\kappa
B_{HV} L}\left({\bf \Pi}^s\right)^{-1}\nabla^2 L,\ee  and assuming
isotropy one gets

\be\label{deTdet5}\pardt T=\frac{1}{\rho c_V}\left[\frac{\rho
\epsilon_L \eta }{\gamma_1\kappa L}-\frac{3\xi}{\kappa B_{HV}
L}\right]\nabla^2 T+\frac{1}{\rho c_V}\left[ \frac{\rho \epsilon_L
\nu }{\gamma_1\kappa L}-\frac{3\chi}{\kappa B_{HV} L}\right]
\nabla^2 L,\ee

From the relations (\ref{deLdet5}) and (\ref{deTdet5}), and from
the positive character of the vortex diffusion coefficient and of
thermal conductivity one deduces  \be \frac{\nu}{\gamma_1}>0 \ee
and \be \frac{3\xi}{\kappa B_{HV} L}-\frac{\rho \epsilon_L \eta
}{\gamma_1\kappa L}<0. \ee

Thus, despite of the high number of coefficients appearing in
(\ref{siste-Ep-qp-Lp-Jp})--(\ref{prod-term-J}), one has been able
to provide a physical interpretation for many of them, which would
allow for their respective measurements in suitable experiments.

\section{Interaction of second sound and vortex density waves}
\setcounter{equation}{0} In this Section wave propagation in
counterflow vortex tangles is studied, with the aim to discuss the
physical effects of the interaction between high-frequency second
sound and vortex waves. Expressing the energy $E$ in terms of $T$
and $L$, the system (\ref{siste-Ep-qp-Lp-Jp}) becomes

\be\label{siste-Tp-qp-Lp-Jp} \left\{
  \begin{array}{ll}
    \rho c_V \pardt T +\rho \epsilon_L \pardt L +\pardxj q_j=0, \\
    \pardt q_i + \xi \pardxi T  +\chi \pardxi L  =\sigma^q_i,\\
    \pardt L +\pardxj J_j  =\sigma^L,\\
    \pardt J_i + \eta \pardxi T  +\nu\pardxi L  =\sigma^{J_i},
  \end{array}
\right. \ee where $c_V= \pard_T E$ is the specific heat at
constant volume and $\epsilon_L=\pard_L E.$ These equations are
analogous to those proposed in \cite{MJ1} except for the choice of
$J_i$: in fact here $J_i$ is assumed to be an independent field
whereas in \cite{MJ1} $J_i$ was assumed as dependent on  $q_i$.
However, at high frequency, $J_i$ will become dominant and will
play a relevant role, as shown in the following.

A stationary solution of the system (\ref{siste-Tp-qp-Lp-Jp}), with
the expressions of the production terms
(\ref{prod-term-q}--\ref{prod-term-J}), is

\be\label{solu-staz-sistema} {\bf q}={\bf q_0}=(q_{01},0,0), \quad
L=L_0=\frac{A^2}{B^2}q_{01}^2, \ee \be T=T_0({\bf
x})=T^*-\frac{\kappa B_{HV}}{\xi} L_0 a q_{01}x_1, \quad {\bf
J}_0=\left(\frac{\kappa B_{HV}}{\xi \gamma_1 \kappa}a
q_{01},0,0\right), \ee with $q_{01}>0$.

The quantities (\ref{prod-term-q}), (\ref{prod-term-L}) and
(\ref{prod-term-J}) can be approximated around the stationary
solutions in the following way

\be\label{prod-term-q-app}
 \sigma_i^{\bf q} \simeq -\frac{1}{2} \kappa B_{HV} \left[(3a-1)\delta_{i1}+(1-a)\right]
\left(q_{i0} (L-L_0)+L_0 q_i\right), \ee
\be\label{prod-term-L-app} \sigma^L\simeq
-\left[2BL_0-\frac{3}{2}AL_0^{1/2}q_{01}\right](L-L_0)+AL_0^{3/2}{\bf
\hat{q}}_0\cdot(\textbf{q}-\textbf{q}_0), \ee and
\be\label{prod-term-J-app} \vec{\sigma}^{J}\simeq -\gamma_1 \kappa
L_0 {\bf J}-\gamma_1 \kappa (L-L_0) {\bf J}_0, \ee where the
subscript $0$ denotes the stationary values for ${\bf q}$, $L$ and
${\bf J}$.

Now, consider the propagation of harmonic plane waves of the four
fields of the equation (\ref{siste-Tp-qp-Lp-Jp}) in the following
form

\be\label{trasf} \left\{
\begin{array}{ll}
T=T_0({\bf x})+\tilde{T} e^{i(K{\bf n\cdot x}-\omega t)}\\
{\bf q}={\bf q}_0+\tilde{{\bf q}} e^{i(K{\bf n\cdot x}-\omega t)}\\
L=L_0+\tilde{L} e^{i(K{\bf n\cdot x}-\omega t)}\\
{\bf J}={\bf J}_0+\tilde{{\bf J}} e^{i(K{\bf n\cdot x}-\omega t)},
\end{array}\right.
\ee where $K=k_r+ik_s$ is the wave number, $\omega$ the real
frequency, ${\bf n}$   the unit vector along the direction of the
wave propagation, and the oversigned quantities denote small
amplitudes of the fields, whose product can be neglected.

Substituting (\ref{trasf}) in the system (\ref{siste-Tp-qp-Lp-Jp}),
the following equations for the  small amplitudes are obtained

\be\label{sistema-linea} \left\{
\begin{array}{ll}
-\omega [\rho c_V]_0\tilde{T}-\omega[\rho\epsilon_L]_0 \tilde{L}+K
\tilde{\bf{q}}\cdot {\bf n}=0 \\
\left[-\omega-\frac{i}{2}\kappa B_{HV}L_0\left((3a-1){\bf c}_1{\bf
c}_1+1-a\right)\right]\tilde{{\bf q}}+ \xi_0 K\tilde{T}{\bf n}\\
\hspace{1.2cm} -\left(-\chi_0 K {\bf n}+ i a\kappa
B_{HV}  {\bf q}_0 \right) \tilde{L}=0\\
\left[-\omega-i\left(2BL_0-\frac{3}{2}AL_0^{1/2}
q_{01}\right)\right] \tilde{L}+K \tilde{{\bf J}}\cdot {\bf n}+i A
L_0^{3/2} \tilde{q}_1=0\\
\left(-\omega -i \gamma_1 \kappa L_0 \right) \tilde{{\bf J}} +
\eta_0 K {\bf n} \tilde{T}+\left(\nu_0 K {\bf n}- i \gamma_1 \kappa
{\bf J}_0 \right) \tilde{L}=0
\end{array}\right.
\ee where ${\bf c}_1$ is their unit vector along the first axis
$x_1$ and ${\bf c}_1{\bf c}_1$ is the dyadic product. Note that
the subscript $0$ refers to the unperturbed state; in what
follows, this subscript will be dropped out to simplify the
notation.

\subsection*{First case: ${\bf n}$ parallel to ${\bf q_0}$}
Now, impose the condition that the direction of the wave
propagation ${\bf n}$ is parallel to the heat flux ${\bf q_0}$,
namely ${\bf n}=(1,0,0)$. Through these conditions the system
(\ref{sistema-linea}) becomes

\be\label{sistema-linea-n//q} \left\{
\begin{array}{ll}
-\omega \rho c_V\tilde{T}+K \tilde{q}_1-\omega \rho\epsilon_L  \tilde{L}=0 \\
\xi K \tilde{T}-\left(\omega+i a\kappa
B_{HV}L\right)\tilde{q}_1-\left(-\chi K +i \kappa
B_{HV} a q_1 \right) \tilde{L}=0\\
i A
L^{3/2} \tilde{q}_1-\left(\omega+i\tau_L^{-1}\right) \tilde{L}+K \tilde{J}_1=0\\
\eta K \tilde{T}+\left(\nu K- i \gamma_1 \kappa J_{1} \right)
\tilde{L}+\left(-\omega -i \gamma_1 \kappa L \right)
\tilde{J}_1=0\\
\\
\left(-\omega-\frac{i}{2}\kappa B_{HV}L(1-a)\right)\tilde{q}_2=0\\
\left(-\omega-\frac{i}{2}\kappa B_{HV}L(1-a)\right)\tilde{q}_3=0\\
\left(-\omega -i \gamma_1 \kappa L\right) \tilde{J}_2 =0\\
\left(-\omega -i \gamma_1 \kappa L \right) \tilde{J}_3=0
\end{array}\right.
\ee where
\[
\tau_L^{-1}=\left(2BL-\frac{3}{2}AL^{1/2} q_{1}\right).
\]
 Note that the transversal modes, those corresponding to the four latter equations,
evolve independently with respect to the longitudinal ones,
corresponding to the four former equations.

One will limit the study to the case in which $\omega$ and the
modulus of the wave number $K$ assume values high enough to make
considerable simplification in the system.  Indeed, it is for high
values of the frequency that the wave behavior of the vortex
tangle can be evidenced because the first term in
(\ref{siste-Tp-qp-Lp-Jp}c) will become relevant, as shown in
Section 2. Note that the assumption $|K|=|k_r+i k_s|$ large refers
to a large value of its real part $k_r$, which is related to the
speed of the vortex wave, whereas the imaginary part $k_s$,
corresponding to the attenuation factor of the wave, will be
assumed small.

This problem is studied into two steps: first  assuming $|K|$ and
$\omega$ extremely high to neglect all terms which do not depend
on them. Then,  the solution so obtained is perturbed in order to
evaluate the influence of the neglected terms on the velocity and
the attenuation of high-frequency waves.

{\bf Step I:} Under the mentioned assumptions the system
(\ref{sistema-linea-n//q}) becomes

\be\label{sistema-linea-n//q-OmegaGrande} \left\{
\begin{array}{ll}
-\omega \rho c_V\tilde{T}+k_r \tilde{q}_1-\omega\rho\epsilon_L \tilde{L}=0 \\
\xi k_r \tilde{T}-\omega\tilde{q}_1+\chi k_r
\tilde{L}=0\\
-\omega\tilde{L}+k_r \tilde{J}_1=0\\
\eta k_r \tilde{T}+\nu k_r \tilde{L}-\omega \tilde{J}_1=0\\
\\
-\omega\tilde{q}_2=0, \quad -\omega\tilde{q}_3=0, \quad -\omega
\tilde{J}_2 =0, \quad -\omega \tilde{J}_3=0
\end{array}\right.
\ee Denoting with $w=\omega/k_r$ the speed of the wave,  the
following dispersion relation is obtained

\be\label{rela-disper-n//q-OmegaGra}
w^4-\left[V_2^2+\nu-\frac{\eta}{\rho c_V}\left(\rho
\epsilon_L-\frac{\chi}{\nu}\right)\right] w^2+ V_2^2 \nu =0, \ee
where $V_2=\left(-\lambda_{11} T^2 \rho c_V\right)^{-1/2}$ is the
second sound speed in the absence of vortex tangle
\cite{MJ1,JLM,M-PRB48} and from (\ref{lam1chi-lam1zet}b) it is
related to the coefficient $\xi$ by the relation $\xi=V_2^2\rho
c_V-\lambda_{12}\eta/\lambda_{11}$. Further, if one assumes that
the coefficient $\eta$ is zero \be\label{eta=0} \eta=0 \qquad
\Rightarrow \qquad \frac{\lambda_{12}}{\lambda_{11}}=T^2{\partial
\over\partial T}\left({\mu^L\over
T}\right)=\frac{2S_3}{S_2}=-\frac{\chi}{\nu}, \ee then the
dispersion relation (\ref{rela-disper-n//q-OmegaGra}) has the
solutions \be\label{rela-disper-n//q-OmegaGra-sempl} w_{1,2}=\pm
V_2, \qquad w_{3,4}=\pm \sqrt{\nu}, \ee to which correspond the
propagation modes shown in Table \ref{Table 1}.

\begin{table}[h]
 \centering
\begin{tabular}{|l|l|}
{$w_{1,2}=\pm V_2$}&{$w_{3,4}=\pm \sqrt{\nu}$}  \\
\hline\hline &                          \\
{$\tilde T=\psi $}&{$\tilde T=-\frac{1}{\rho
c_V}\left(\frac{\chi-\nu\rho\epsilon_L}{V_2^2-\nu}\right)\psi$}\\
{$\tilde q_1=\pm V_2 \rho c_V\psi$}&{$\tilde q_1=\pm
\frac{\sqrt{\nu} \left(\rho\epsilon_L V_2^2-\chi\right)}{V_2^2-\nu}  \psi$}  \\
{$\tilde L=0  $ } &{$\tilde L=\psi $}\\
{$\tilde J_1=0  $ } &{$\tilde J_1=\pm \sqrt{\nu} \psi$}\\
 \end{tabular}
   \caption{\small{Modes corresponding to second sound velocity and vortex waves, respectively.}} \label{Table 1}
\end{table}

As one sees from the first column of Table \ref{Table 1}, under
the hypothesis (\ref{eta=0}) the high-frequency wave of velocity
$w_{1,2}=\pm V_2$ is a temperature wave (\textrm{i.e.} the second
sound) in which the two quantities $\tilde L$ and $\tilde J_1$ are
zero, whereas in the second column the high-frequency wave of
velocity $w_{3,4}=\pm \sqrt{\nu}$ is a wave in which all fields
vibrate. The latter result is logic because when the vortex wave
is propagated in the superfluid helium, temperature $T$ and heat
flux $q_1$ cannot remain constant. This behavior is different from
that obtained in \cite{JMS-Ph.Lett.A}, because using that model in
the second sound also the line density $L$ vibrates. In fact,
there the flux of vortices ${\bf J}$ was chosen proportional to
${\bf q}$, so that vibrations in the heat flux (second sound)
produce vibrations in the vortex tangle. Experiments on
high-frequency second sound are needed to confirm this new result.

{\bf Step II:} Suppose that the terms of the system
(\ref{sistema-linea-n//q}), which don't appear in the system
(\ref{sistema-linea-n//q-OmegaGrande}), and the term $\eta$ are
small enough to be considered as perturbations of the velocity $w$
of the wave and of the attenuation term $k_s$ of the wave number
$K.$ Substituting the following assumptions
\[
\bar{w}=\frac{\omega}{k_r}=w+\delta \quad \textrm{and} \quad
K=k_r+ik_s
\]
in the system (\ref{sistema-linea-n//q}), one find the expression
(\ref{rela-disper-n//q-OmegaGra-sempl}), at the zeroth order in
$\delta$ and $k_s$, whereas at the first order in $\delta$ and
$k_s$, one obtains

\be\label{w12-pert-par}
  \bar{w}_{1,2} = \left(1-\frac{\eta}{2\rho c_V \left(w_{1,2}^2-
w_{3,4}^2\right)} \left(\rho \epsilon_L
-\frac{\chi}{w_{3,4}^2}\right)\right)w_{1,2},\ee
\be\label{w34-pert-par} \bar{w}_{3,4}  = \left(1+\frac{\eta}{2\rho
c_V \left(w_{1,2}^2- w_{3,4}^2\right)} \left(\rho \epsilon_L
-\frac{\chi}{w_{3,4}^2}\right)\right)w_{3,4}, \ee and
\be\label{ks12-parall}
  k_s^{(1,2)}  = \frac{a \kappa L
B_{HV}}{2w_{1,2}}+ \frac{A L^{3/2} \left(w_{1,2}^2 \rho
\epsilon_L- \chi\right)}{2\left(w_{1,2}^2- w_{3,4}^2\right)}, \ee
\be\label{ks34-parall} k_s^{(3,4)} = \frac{\kappa L
\gamma_1+\tau_L^{-1}}{2w_{3,4}}-\frac{A L^{3/2} \left(w_{1,2}^2
\rho \epsilon_L-\chi\right)}{2\left(w_{1,2}^2-
w_{3,4}^2\right)}+\frac{J_1\kappa \gamma_1}{2 w_{3,4}^2}. \ee
Observe that in this approximation all thermodynamical fields
vibrate simultaneously and the attenuation coefficients $k_s$ are
influenced by the choice of $\bf J$ as independent variable, as
one easily sees by comparing expressions
(\ref{ks12-parall}--\ref{ks34-parall}) with those obtained in
\cite{JMS-Ph.Lett.A}. Looking at these results, in particular the
two speeds (\ref{w12-pert-par}--\ref{w34-pert-par}), one sees that
these velocities are not modified when one makes the simplified
hypothesis that the coefficient $\eta$ is equal to zero. In
\cite{JMS-Ph.Lett.A} it was observed that the second sound
velocity is much higher than that of the vortex waves, so that the
small quantity $\eta$ should influence the two velocities
(\ref{w12-pert-par}-\ref{w34-pert-par}) in a different way:
negligible for the second sound velocity but relevant for the
vortex waves. Regarding the attenuation coefficients
(\ref{ks12-parall}--\ref{ks34-parall}), one sees that the first
term in (\ref{ks12-parall}) is identical to that obtained in
\cite{JLM}, when the vortices are considered fixed. The new term,
proportional to $A$, comes from the interaction between second
sound and vortex waves.

It is to note that the first term (\ref{ks12-parall}) produces an
attenuation both to forward waves and to backward waves, while the
second term contributes to the two kinds of waves in a opposite
way, according to the sign of this term. Detailed measurements of
the attenuation of second sound in directions parallel and
orthogonal to the heat flux could allow us to establish the
presence and the sign of this term.

Note also that the second term of the dissipative coefficient
$k_s^{(1,2)}$ is the same as the third term of $k_s^{(3,4)}$, but
with an opposite sign. This means that this term contributes to
the attenuation of the two waves in opposite ways; and its
contribution depends also on whether the propagation of forward
waves or of backward waves is considered. The first term of
$k_s^{(3,4)}$ produces always an attenuation of the wave, while
the behavior of the third term is analogous to the first one.

\subsection*{Second case: ${\bf n}$ orthogonal to ${\bf q_0}$}
In order to make a more detailed comparison with the model studied
in \cite{MJ1,JMS-Ph.Lett.A}, one proceeds to analyze another
situation, in which the direction of the wave propagation is
perpendicular to the heat flux, that is, for example, assuming
${\bf n}=(0,0,1)$. This choice simplifies the
system~(\ref{sistema-linea}) in the following form

\be\label{sistema-linea-n-per-q} \left\{
\begin{array}{ll}
-\omega \rho c_V\tilde{T}+K
\tilde{q}_3-\omega\rho\epsilon_L \tilde{L}=0 \\
\left(-\omega-i\kappa B_{HV}L a\right)\tilde{q}_1 -i \kappa
B_{HV} a q_{1}\tilde{L}=0\\

\xi K
\tilde{T}-\left(\omega+\frac{i}{2}\kappa B_{HV}L(1-a)\right)\tilde{q}_3+\chi K \tilde{L}=0\\
i A
L^{3/2} \tilde{q}_1-\left(\omega+i\tau_L^{-1}\right) \tilde{L}+K \tilde{J}_3=0\\
\eta K\tilde{T}+\nu K \tilde{L}+ \left(-\omega -i \gamma_1 \kappa
L\right) \tilde{J}_3 =0\\
\\
\left(-\omega-\frac{i}{2}\kappa B_{HV}L (1-a)\right)\tilde{q}_2=0\\
-i \gamma_1 \kappa J_{1} \tilde{L}+\left(-\omega -i \gamma_1 \kappa
L \right)
\tilde{J}_1=0\\
\left(-\omega -i \gamma_1 \kappa L \right) \tilde{J}_2=0
\end{array}\right.
\ee Note that, in contrast with what was seen before, but in
agreement with the corresponding situation of the model described
in \cite{MJ1,JMS-Ph.Lett.A}, here the transversal and the
longitudinal modes in general do not evolve independently, as
shown from the first five equations. However, one will see that
this is the case if high-frequency waves are considered.

As in the previous situation, assume that the values of the
frequencies $\omega$ and of the real part of the wave number,
$k_r$, are high enough, such that the system
(\ref{sistema-linea-n-per-q}) may be simplified in the following
form

\be\label{sistema-linea-n-per-q-OmegaGra} \left\{
\begin{array}{ll}
-\omega \rho c_V\tilde{T}+k_r
\tilde{q}_3-\omega\rho\epsilon_L \tilde{L}=0 \\
-\omega\tilde{q}_1 =0\\
\xi k_r
\tilde{T}-\omega\tilde{q}_3+\chi k_r \tilde{L}=0\\
-\omega \tilde{L}+k_r \tilde{J}_3=0\\
\eta k_r\tilde{T}+ \nu k_r  \tilde{L} -\omega  \tilde{J}_3 =0\\
\\
-\omega\tilde{q}_2=0 \quad -\omega\tilde{J}_1=0 \quad -\omega
\tilde{J}_2=0
\end{array}\right.
\ee Note that in this special case, as in the previous case and in
\cite{JMS-Ph.Lett.A}, only the longitudinal modes are present, so
that the dispersion relation assumes the form

\be\label{rela-disper-nORTq-OmegaGra}
w\left(w^4-\left[V_2^2+\nu-\frac{\eta}{\rho c_V}\left(\rho
\epsilon_L+\frac{\lambda_{12}}{\lambda_{11}}\right)\right] w^2+
V_2^2 \nu\right) =0, \ee which is similar to equation
(\ref{rela-disper-n//q-OmegaGra}).

Now, the arguments suggested are the same to the previous
situation, in fact, under the hypothesis (\ref{eta=0}), the
dispersion relation (\ref{rela-disper-nORTq-OmegaGra}) takes the
form \be\label{rela-disper-n-orto-q-OmegaGra-sem} w
(w^2-\nu)(w^2-V_2^2) =0, \ee where $V_2$ is the second sound
velocity and $\sqrt{\nu}$ is the velocity of the vortex density
waves in helium II. The conclusions which one achieves here are
the same to those of the previous situation. Indeed, the modes
corresponding to the solutions
(\ref{rela-disper-n-orto-q-OmegaGra-sem}) are showed in Table
\ref{Table 2}, which, apart from the first column, are identical
to those shown in Table \ref{Table 1}.

\begin{table}[h]
 \centering
\begin{tabular}{|l|l|l|}
{$w_{0}=0$}&{$w_{1,2}=\pm V_2$}&{$w_{3,4}=\pm \sqrt{\nu}$}  \\
\hline\hline &                          \\
{$\tilde q_1=\psi $}&{$\tilde q_1=0 $}&{$\tilde q_1=0$}\\
{$\tilde T=0 $}&{$\tilde T=\psi $}&{$\tilde T=-\frac{1}{\rho c_V}\left(\frac{\chi-\nu\rho\epsilon_L}{V_2^2-\nu}\right)\psi$}\\
{$\tilde q_3=0$}&{$\tilde q_3=\pm V_2 \rho c_V\psi$}&{$\tilde
q_3=\pm \frac{\sqrt{\nu} \left(\rho \epsilon_L V_2^2-\chi\right)}{V_2^2-\nu}\psi$}  \\
{$\tilde L=0  $ } &{$\tilde L=0  $ } &{$\tilde L=\psi $}\\
{$\tilde J_3=0  $ } &{$\tilde J_3=0  $ } &{$\tilde J_3=\pm \sqrt{\nu} \psi$}\\
\end{tabular}
   \caption{\small{Modes corresponding to null velocity, second sound velocity, and vortex waves, respectively.}}
   \label{Table 2}
\end{table}

Now,  the same procedure than in  the previous situation is
followed, that is one supposes that all the quantities of the
system (\ref{sistema-linea-n-per-q}), which don't appear in the
system (\ref{sistema-linea-n-per-q-OmegaGra}), are small enough
compared to the other terms of the same system. Further, one also
assume that the coefficient $\eta$ is not zero, but it has small
enough values to be considered as a small perturbation to the
physical system. Therefore, one assumes
\[
\bar{w}=\frac{\omega}{k_r}=w+\delta \quad \textrm{and} \quad
K=k_r+ik_s,
\]
and substituting them in the dispersion relation of the system
(\ref{sistema-linea-n-per-q}), one finds the relation
(\ref{rela-disper-n-orto-q-OmegaGra-sem}), at the zeroth order in
$\delta$ and $k_s$, and the following two expressions at the first
order in $\delta$ and $k_s$

\be\label{w12-pert-or}
  \bar{w}_{1,2} = \left(1-\frac{\eta}{2\rho c_V \left(w_{1,2}^2-
w_{3,4}^2\right)} \left(\rho \epsilon_L
-\frac{\chi}{w_{3,4}^2}\right)\right)w_{1,2}, \ee
\be\label{w34-pert-or}\bar{w}_{3,4} =\left(1+\frac{\eta}{2\rho c_V
\left(w_{1,2}^2- w_{3,4}^2\right)} \left(\rho \epsilon_L
-\frac{\chi}{w_{3,4}^2}\right)\right)w_{3,4}, \ee
 and
\be\label{ks12-ort}
  k_s^{(1,2)} = \frac{(1-a)\kappa L B_{HV}}{4 w_{1,2}},\ee
\be\label{ks34-ort}k_s^{(3,4)} =  \frac{\tau_L^{-1}+\kappa L
\gamma_1}{2 w_{3,4}}. \ee As regards the expression
(\ref{ks12-ort}) for the dissipative term $k_s^{(1,2)}$, note that
it is the same as the expression obtained when the vortices are
assumed fixed \cite{JM-PRB74, PS-PB2007}, whereas the attenuation
term $k_s^{(3,4)}$ is the same as the second term of $k_s^{(3,4)}$
of the first case (${\bf n}$ parallel to ${\bf q_0}$).

\section{Conclusions}
\par\noindent
The previous hydrodynamical model of inhomogeneous turbulent
vortex tangle \cite{MJ1,JMS-Ph.Lett.A} has been generalized, by
adding the vortex flux ${\bf J}$ to the set of the independent
variables $E$, ${\bf q}$ and $L$. In this new model, ${\bf J}$ is
no longer described by a usual constitutive equation but it has
its own dynamical equation.

A set of evolution equations for $E$, ${\bf q}$, $L$, and ${\bf
J}$ subject to the restrictions of the second law of
thermodynamics are studied and used to analyze the behavior of
second sound and vortex waves, with special emphasis on their
mutual couplings. This mathematical analysis may be useful in the
interpretation of high-frequency second sound experiments,  which
play a key role as a tool for the probe of small spatial scales of
turbulent vortex tangles. Indeed, the time derivative of ${\bf J}$
becomes relevant at high enough frequencies; in such a regime, the
behavior of vortices becomes undulatory instead of being diffusive
--- the behavior assumed in the previous studies
\cite{MJ1,JMS-Ph.Lett.A}.

An interesting result is found from the comparison between the
results of the wave propagation parallel and orthogonal to the
heat flux. In fact, when the waves propagate orthogonal to the
heat flux, the presence of the vortex tangle always causes  an
attenuation of the waves. But, when the propagation of the wave is
collinear to the heat flux other terms are present. These terms
have a positive or negative contribution depending on whether the
direction of the wave is the same or opposite to the direction of
the heat flux.

Now, compare the results of the perturbed situation of the first
case, when ${\bf n}$ is parallel to ${\bf q}$, with those obtained
in \cite{MJ1,JMS-Ph.Lett.A}. The results of the comparison
regarding the second case, ${\bf n}$ normal to ${\bf q}$, are the
same to those of the first case. As in \cite{MJ1,JMS-Ph.Lett.A},
in this case one has the propagation of two kinds of waves, namely
heat waves and vortex waves, which cannot be considered as
propagating independently from each other. In fact, the uncoupled
situation (equation (\ref{rela-disper-n//q-OmegaGra-sempl})), in
which the propagation of the second sound is not influenced by the
fluctuations of the vortices, is no more the case when the
quantities $N_1=a\kappa B_{HV}L$, $N_2=\kappa B_{HV} a q_1,$
$N_3=A L^{3/2},$ $N_4=\gamma_1 \kappa J_{1},$ $N_5=\gamma_1 \kappa
L,$ $\tau_L^{-1}$ and $\eta$, appearing in the system
(\ref{sistema-linea-n//q}), are considered. Indeed, from
(\ref{w12-pert-par}--\ref{w34-pert-par}) and from the results of
\cite{JMS-Ph.Lett.A} one makes in evidence that heat and vortex
waves cannot be considered separately, that is as two different
waves, but as two different features of the same phenomena. Of
course, the results obtained here are more exhaustive than those
of \cite{MJ1,JMS-Ph.Lett.A}: in fact, comparing the velocities at
the first order of approximation in both models, one deduces  that
the expressions (\ref{w12-pert-par}--\ref{w34-pert-par}) depend
not only on the velocities of heat waves and vortex waves, as in
\cite{MJ1,JMS-Ph.Lett.A}, but also on the coefficient $\eta$,
which comes from the equation (\ref{siste-Tp-qp-Lp-Jp}d) of the
vortex flux $\bf J$, and whose physical meaning is a
thermodiffusion coefficient of vortices. The fourth equation of
the system (\ref{sistema-linea-n//q}) shows that the vortex flux
$\tilde J_1$ is not proportional to the heat flux, as it was
assumed in \cite{MJ1,JMS-Ph.Lett.A}, but it satisfies an equation
in which also the fields $\tilde L$ and $\tilde T$, through
$\eta$, are present.

It is to note that the attenuation of the second sound depends on
the relative direction of the wave with respect to the heat flux:
in some experiments this dependence was shown for parallel and
orthogonal directions \cite{AMS-PRL53}. These results were
explained assuming an anisotropy of the tangle of vortices, which
corresponds to the presence of the parameter $a$ in
(\ref{tensorPi}). But, looking at the expressions
(\ref{ks12-parall}) and (\ref{ks12-ort}) of the attenuation of the
second sound in the high-frequency regime, one notes that these
expressions are not equal after assuming $a=1/3$ (isotropy of the
tangle). In particular, the term  \be \frac{A L^{3/2}
\left(w_{1,2}^2 \rho \epsilon_L- \chi\right)}{2\left(w_{1,2}^2-
w_{3,4}^2\right)}\ee in (\ref{ks12-parall}) causes a dependence of
the attenuation depending on whether the wave direction agrees
with the direction of the heat flux ${\bf q}$ or not. This term is
absent if the wave propagates orthogonal to the heat flux. Note,
in contrast, that the propagation speeds (\ref{w12-pert-par}) and
(\ref{w34-pert-par}) for propagation direction ${\bf n}$ parallel
to ${\bf q}$ coincide with (\ref{w12-pert-or}) and
(\ref{w34-pert-or}), respectively, for propagation direction ${\bf
n}$ normal to ${\bf q}$. Thus, the behavior of speed of
propagation is isotropic and does not depend on the isotropy or
anisotropy of the tangle.

In conclusion, it could be that an anisotropy of the behavior of
high-frequency second sound does not necessarily imply an actual
anisotropy of the tangle in pure counterflow regime, but only a
different behavior of the second sound due to the interaction with
the vortex waves. This may be of interest if one wants to explore
the degree of isotropy at small spatial scales. Of course, some
more experiments are needed in order to establish the presence and
the sign of these additional terms.

\section*{Acknowledgements}
The authors acknowledge the support of the Acci\'{o}n Integrada
Espa\~{n}a-Italia (Grant S2800082F HI2004-0316 of the Spanish
Ministry of Science and Technology and grant IT2253 of the Italian
MIUR) and the collaboration agreement between Universit\`{a} di
Palermo and Universitat Aut\`{o}noma de Barcelona. DJ acknowledges
the financial support from the Direcci\'{o}n General de
Investigaci\'{o}n of the Spanish Ministry of Education under grant
Fis2006-12296-c02-01 and of the Direcci\'{o} General de Recerca of
the Generalitat of Catalonia, under grant 2005 SGR-00087. MSM and
MS acknowledge the financial support "Fondi 60\%" of the
Universit\`{a} di Palermo. MS acknowledges the "Assegno di
ricerca" of the Universit\`{a} di Palermo.

\end{document}